\documentstyle[aps,12pt,epsfig,manuscript]{revtex}
\global\firstfigfalse
\begin{document}
\preprint{INJE-TP-99-4}
\def\overlay#1#2{\setbox0=\hbox{#1}\setbox1=\hbox to \wd0{\hss #2\hss}#1%
\hskip -2\wd0\copy1}

\title{Tachyon and fixed scalars of D$5_{\pm}$-D$1_{\pm}$ black hole 
in type 0B string theory}

\author{ H.W. Lee$^1$, Y.S. Myung$^1$ and  Jin Young Kim$^2$ }
\address{$^1$Department of Physics, Inje University, Kimhae 621-749, Korea\\
$^2$ Department of Physics, Kunsan National University, Kunsan 573-701, Korea}

\maketitle

\begin{abstract}
In the type 0B string theory, 
we discuss the role of tachyon($T$) and fixed scalars($\nu,\lambda$).
The issue is to 
explain the difference between tachyon and fixed scalars
in the D$5_{\pm}$-D$1_{\pm}$ black hole background. 
For this purpose, we perform the semiclassical calculation.
Here one finds a mixing between ($\nu, \lambda, T$) and the other fields.
Using the decoupling procedure, one finds the linearized equation 
for the tachyon. From the potential analysis, it turns out that 
$\nu$ plays a role of test field well, while the tachyon 
induces an instability of Minkowski space vacuum.
But the roles of $\nu$ and $T$ are the same in the near-horizon geometry.
Finally we discuss the stability problem.

\end{abstract}
\vfill
Compiled at \today : \number \time.

\newpage
\section{Introduction}
\label{introduction}
Recently type 0 string theories attract much interest
in the study of non-supersymmetric gauge theories
\cite{Pol99IJMPA645,Kle99NPB155,Min99JHEP01020}.
Type 0 string theories can be obtained from the worldsheet of 
type II string theories by performing a non-chiral GSO projection
\cite{Dix86NPB93}.
The resulting theories have world sheet supersymmetry but no 
space-time supersymmetry.
The crucial differences of type 0 theories with type II theories 
is to have the doubling of Ramond-Ramond(RR) fields and the tachyon. 
Thus they all have twice as many D-branes.
The application of this model to the study of non-SUSY gauge theory 
was realized by Klebanov and Tseytlin\cite{Kle99NPB155}.
They considered the theory of N coincident electrically charged 
D3-branes. Using the near-horizon D3-brane, they 
constructed an SU(N) gauge theory with six adjoint scalars and 
studied its behavior. They observed that the doubled RR flux 
in type 0 dual background stabilizes the tachyon.
Since then a number of papers on this mode appeared
\cite{Fer9811208}.
In a subsequent paper, Klebanov and Tseytlin\cite{Kle99JHEP03015} also 
considered the theory of N electric D$3_{+}$-branes coincident with N
magnetic D3$_-$-branes. The non-SUSY theory is the 
SU(N)$\times$SU(N) gauge theory coupled to six adjoint scalars 
of the first SU(N), six adjoint scalars of the second SU(N), and 
fermions in the bifundamental representation(four Weyl spinors in 
the (N,$\bar{\rm N}$) and another four in the ($\bar{\rm N}$,N)).
The D$p_{\pm}$-brane bound  states of the type 0 theories are somewhat 
similar to those of the type II theories.
However, the essential difference with type II theories comes from 
the existence of tachyon.
In the previous work of one of the authors\cite{Kim9906196},
the low energy scattering of fields in the type 0B theory for electric
D$3$-branes attepmted. In this analysis
the dilaton field can be used as a test field.

One of the simplest way to see the role of tachyon in type 0 theories is
to consider the intersecting D$p$-branes.
The D$p_{\pm}$-brane bound states 
can be intersected according to the same rules of the type II theories.
The D$5_{\pm}$-D$1_{\pm}$ brane black hole thus is constructed 
to show that the corresponding near-horizon geometry is 
AdS$_3\times$S$^3\times$T$^4$ and it has asymptotically flat 
space at infinity.
It is shown that the tachyon field can be stabilized in 
AdS$_3 \times$S$^3 \times$T$^4$\cite{Kle99JHEP03015,Cos9903128}. 
In this paper, we will study the role of the tachyon in 
D5$_\pm$-D1$_\pm$ brane black hole, by comparing 
it with the fixed scalars.
Further, we will present a complete solution to the stability by 
analyzing the potentials surrounding the D5$_\pm$-D1$_\pm$ 
brane black hole.

The organization of the paper is as follows.
In section \ref{blackhole}, we briefly sketch the D5$_\pm$-D1$_\pm$ brane 
black hole in the type 0B string theory.
We set up the perturbation for all fields around this black hole 
background in Sec. \ref{perturbation}.
Here we choose the harmonic gauge and use all linearized equations 
to decouple ($\nu,t$) from the remaining fields.
Sec. \ref{potential} is devoted to analyzing their potentials.
Finally we discuss our results in Sec. \ref{discussion}.

\section{D$5_{\pm}$-D$1_{\pm}$ brane Black Hole}
\label{blackhole}
Here we consider a class of 5D black holes representing 
the bound state of the D5$_\pm$-D1$_\pm$ brane system compactified on 
T$^5(=$T$^4 \times $S$^1$).  
This black hole can also be obtained as a solution 
to the semiclassical action of type 0B string 
compactified on T$^5$.
The effective action for a 5D black hole 
is given by\cite{Cos9903128}
\begin{eqnarray}
S&=& {1 \over {2 \kappa_5^2}} \int d^5 x \sqrt{-g} \left \{ R - 
{4 \over 3} \left (\nabla \lambda\right )^2 - 4 \left (\nabla \nu \right )^2 
-{1 \over 4} \left (\nabla T \right )^2 -{m^2 \over 4} e^{-2 \lambda/3} T^2 
\right .
\nonumber \\
&& \qquad \qquad
-{1 \over 4} e^{{8 \over 3} \lambda} F^{(K)2} 
-{1 \over 4} e^{-{4 \over 3} \lambda + 4 \nu} 
\left ( f_+(T) F_+^2 +  f_-(T) F_-^2 \right )
\nonumber \\
&& \qquad \qquad
-{1 \over 4} e^{-{4 \over 3} \lambda -4 \nu} 
\left ( f_+(T) H_+^2 +  f_-(T) H_-^2 \right )
\Bigg \} ,
\label{eq-action}
\end{eqnarray}
where $F^{(K)}_{\mu\nu}$ is the Kaluza-Klein(KK) field strength along the 
string direction(S$^1$), $F_{\pm\mu\nu}$ are the electric components of 
the Ramond-Ramond(RR) three-form $F_{3\pm}$ and $H_{\pm\mu\nu}$ are dual to 
the magnetic components of the RR 
three-form $F_{3\pm}$.  Here we omit the analysis of the 6D dilaton $\phi_6$, 
since it is just a minimally 
decoupled scalar.  On the other hand, the scalars $\nu$ and $\lambda$ 
interact with the gauge fields and are examples of the fixed scalar.  $\nu$ is
related to the scale of the internal torus(T$^4$), 
while $\lambda$ is related to 
the scale of the KK circle(S$^1$).  
$f_\pm(T) = 1 \pm T + {\cal O}(T^2)$ and the tachyon mass is given by 
$m^2 = -2/\alpha'$. Comparing with the results of type IIB theory
\cite{Cal97NPB65,Lee98PRD104006}, the new ingredients are the 
tachyon and the doubling of the RR fields.
$\kappa_5^2$ is the 5D  gravitational 
coupling constant ($\kappa_5^2=8 \pi G_N^5, G_N^5$=5D Newtonian 
constant).  This can be determined by $G_N^5 = {G_N^{10} \over V_5} =
{8 \pi^6 g^2 \over (2\pi)^5 VR} = {\pi g^2 \over 4 VR}$ 
with $V=R_5R_6R_7R_8$(volume of $T^4$), 
$R=R_9$(radius of $S^1$), $\alpha'=1,$ and 
$g$(=10D string coupling constant).  
We wish to follow the MTW conventions\cite{Mis73}.

The equations of motion for action (\ref{eq-action}) are given by
\begin{eqnarray}
&&R_{\mu \nu} - {4 \over 3} \partial_\mu \lambda \partial_\nu \lambda 
- 4 \partial_\mu \partial_\nu \nu 
-{1 \over 4} \partial_\mu T \partial_\nu T 
-{1 \over 12} m^2 e^{-2 \lambda/3} T^2 g_{\mu\nu}
\nonumber \\
&&
- e^{{8 \over 3} \lambda} \left ( {1 \over 2} F^{(K)}_{\mu \rho}
  F^{(K)~\rho}_{~~~~\nu} - {1 \over 12} F^{(K)2} g_{\mu \nu} \right )
\nonumber \\
&&
- e^{-{4 \over 3} \lambda + 4 \nu} 
\left \{ 
f_+(T) \left ( 
{1 \over 2}F_{+\mu \rho} F_{+\nu}^{~~\rho} 
- {1 \over 12} F_+^2 g_{\mu \nu} \right ) 
+ f_-(T) \left ( 
{1 \over 2}F_{-\mu \rho} F_{-\nu}^{~~\rho} 
- {1 \over 12} F_-^2 g_{\mu \nu} \right ) 
\right \}
\nonumber \\
&&
- e^{-{4 \over 3}\lambda -4 \nu} 
\left \{ 
f_+(T) \left ( 
{1 \over 2}H_{+\mu \rho} H_{+\nu}^{~~\rho} 
- {1 \over 12} H_+^2 g_{\mu \nu} \right ) 
+ f_-(T) \left ( 
{1 \over 2}H_{-\mu \rho} H_{-\nu}^{~~\rho} 
- {1 \over 12} H_-^2 g_{\mu \nu} \right ) 
\right \}
= 0,
\label{eq-ricci} \\
&& 8 \nabla^2 \nu - e^{-{4 \over 3} \lambda+4 \nu} 
\left ( f_+(T) F_+^2 + f_-(T) F_-^2 \right ) 
+ e^{-{4 \over 3}\lambda -4 \nu} 
\left ( f_+(T) H_+^2 + f_-(T) H_-^2 \right ) = 0 \label{eq-nu}\\
&& 8  \nabla^2 \lambda - 2 e^{{8 \over 3} \lambda} F^{(K)2}
+ e^{-{4 \over 3} \lambda +4 \nu} 
\left ( f_+(T) F_+^2 + f_-(T) F_-^2 \right )
\nonumber \\
&&\qquad \qquad
+ e^{-{4 \over 3} \lambda -4 \nu} 
\left ( f_+(T) H_+^2 + f_-(T) H_-^2 \right )
+ {m^2 \over 2} e^{-2 \lambda/3} T^2 
= 0, \label{eq-lambda}\\
&&\nabla_\mu \left ( e^{{8 \over 3} \lambda} F^{(K)\mu \nu} \right ) =0, 
   \label{eq-fk} \\
&&\nabla_\mu \left ( f_+(T) e^{-{4 \over 3} \lambda+4\nu} F_+^{\mu \nu} \right ) =0, 
   \label{eq-fp} \\
&&\nabla_\mu \left ( f_-(T) e^{-{4 \over 3} \lambda+4\nu} F_-^{\mu \nu} \right ) =0, 
   \label{eq-fm} \\
&&\nabla_\mu \left ( f_+(T) e^{-{4 \over 3} \lambda-4\nu} H_+^{\mu \nu} \right ) =0, 
   \label{eq-hp} \\
&&\nabla_\mu \left ( f_-(T) e^{-{4 \over 3} \lambda-4\nu} H_-^{\mu \nu} \right ) =0, 
   \label{eq-hm} \\
&& \nabla^2 T - m^2 e^{-2 \lambda/3} T 
-{1 \over 2} e^{-4 \lambda/3 + 4 \nu} 
\left ( f'_+(T) F_+^2 + f'_-(T) F_-^2 \right )
\nonumber \\
&&\qquad \qquad \qquad \qquad
-{1 \over 2} e^{-4 \lambda/3 - 4 \nu} 
\left ( f'_+(T) H_+^2 + f'_-(T) H_-^2 \right )
=0,
\label{eq-t}
\end{eqnarray}
where the prime($'$) denotes the differentiation with respect to its argument.
In addition, we need the remaining Maxwell equations as five 
Bianchi identities\cite{Lee98PRD104006} 
\begin{equation}
\partial_{[\mu}F^{(K)}_{~~~~\rho\sigma]}=
\partial_{[\mu}F_{\pm\rho\sigma]}=\partial_{[\mu}H_{\pm\rho\sigma]}=0.
\label{bianchi}
\end{equation}

The black hole solution is given by the background metric  
\begin{equation}
ds^2 = - d f^{-{2 \over 3}}dt^2 +d^{-1}f^{1 \over 3} dr^2 + 
r^2 f^{1 \over 3}d \Omega^2_3
\label{defg}
\end{equation}
and 
\begin{eqnarray}
&&e^{2\bar \lambda} = { f_K \over \sqrt{f_1 f_5}},~~~~
e^{4\bar \nu} = { f_1 \over f_5}, ~~~~f=f_1 f_5 f_K, \label{sol-scalar} \\
&&\bar F^{(K)}_{t r} = { 2 \tilde Q_K \over {r^3 f_K^2}}, ~~~~
\bar F_{\pm t r} = { 2 Q_{1\pm} \over {r^3 f_1^2}}, ~~~~
\bar H_{\pm t r} = { 2 Q_{5\pm} \over {r^3 f_5^2}}, ~~~~
\bar T = 0. \label{sol-vector}
\end{eqnarray}
Here four harmonic functions are defined by
\begin{equation}
f_1= 1 +{ r_1^2 \over r^2}, ~~~~ f_5= 1 +{ r_5^2 \over r^2},~~~~
f_K = 1 + { r_K^2 \over r^2},~~~~ d = 1 - { r_0^2 \over r^2},
\label{harmonics}
\end{equation}
with $r_i^2=r_0^2 \sinh^2\sigma_i,~ i=1,5,K$.
$Q_{1\pm}$, $Q_{5\pm}$ and $\tilde Q_K$ are related to 
the characteristic radii $r_1$, $r_5$, 
$r_K$ and the radius of horizon $r_0$ as
\begin{eqnarray}
&&Q_{j+}= Q_{j-} \equiv Q_j, j=1,2 , ~
\tilde Q_i = {1 \over 2} r_0^2 \sinh 2 \sigma_i, i=1,2,K,
\nonumber \\
&&\tilde Q_j^2 \equiv 2 Q_j^2 = 
Q_{j+}^2 + Q_{j-}^2 = r_j^2 (r_j^2 + r_0^2), j=1,2, ~ 
\tilde Q_K^2 = r_K^2 (r_K^2 + r_0^2), 
\nonumber \\
&& r_i^2 = \sqrt{\tilde Q_i^2 + {r_0^4 \over 4}} - 
{r_0^2 \over 2}, i=1,2,K. \label{eq-charges}
\end{eqnarray}

The background metric (\ref{defg}) is just a 5D Schwarzschild one 
with time and space components rescaled by different powers of $f$.  The event 
horizon (outer horizon) is clearly at $r=r_0$.  When all five charges are 
nonzero, the surface of $r=0$ becomes a smooth inner horizon (Cauchy horizon).  
If one of the charges is zero, the surface of $r=0$ becomes singular.  The 
extremal case corresponds  to the limit of $r_0 \rightarrow 0$ with the boost 
parameters $\sigma_i \rightarrow \pm \infty$ keeping 
$\tilde Q_i$ fixed.  Here one has $\tilde Q_1 = r_1^2,~\tilde Q_5 = r_5^2$, 
and $\tilde Q_K=r_K^2$.  
In this work we are very interested in the limit 
of $r_0, r_K \ll r_1, r_5$, which is called 
the dilute gas approximation.  This corresponds to the near-extremal black hole 
and its thermodynamic quantities are given by
\begin{eqnarray}
&&M_{next} = {2 \pi^2 \over \kappa_5^2} \left ( r_1^2 +
r_5^2 + {1 \over 2}r_0^2\cosh 2 \sigma_K \right ),\label{next-mass}\\
&&S_{next} = { 4 \pi^3 r_0 \over \kappa_5^2} r_1 r_5\cosh \sigma_K
, \label{next-entropy}\\
&&{1 \over T_{H,next}} = {2 \pi \over r_0} r_1 r_5 \cosh \sigma_K.
\label{next-hawking}
\end{eqnarray}
The above energy and entropy are actually those of a gas of 
massless 1D particles.  
In this case the effective temperatures of the left and right moving 
string modes are given by
\begin{equation}
T_L = {1 \over 2 \pi} \left ( {r_0 \over r_1 r_5} \right ) e^{\sigma_K},~~
T_R = {1 \over 2 \pi} \left ( {r_0 \over r_1 r_5} \right ) e^{-\sigma_K}.
\label{eff-temp}
\end{equation}
The Hawking temperature is given by their harmonic average
\begin{equation}
{2 \over T_H} = {1 \over T_L} + {1 \over T_R}.
\label{eff-hawking}
\end{equation}

\section{Perturbation Analysis}
\label{perturbation}
Here we start with the perturbation 
around the black hole background as in \cite{Lee98PRD104006}
\begin{eqnarray}
&&F^{(K)}_{tr} = \bar F^{(K)}_{tr} + {\cal F}^{(K)}_{tr} = 
\bar F^{(K)}_{tr} [1 + {\cal F}^{(K)}(t,r,\chi, \theta,\phi)],
\label{ptrFk} \\
&&F_{\pm tr} = \bar F_{\pm tr} + {\cal F}_{\pm tr} = 
\bar F_{\pm tr} [1 + {\cal F}_\pm(t,r,\chi,\theta,\phi)],
\label{ptrF} \\
&&H_{\pm tr} = \bar H_{\pm tr} + {\cal H}_{\pm tr} = 
\bar H_{\pm tr} [1 + {\cal H}_\pm(t,r,\chi,\theta,\phi)],
\label{ptrH} \\    
&&\lambda = \bar \lambda + \delta\lambda(t,r,\chi,\theta,\phi),  
\label{ptr-lambda}  \\  
&&\nu = \bar \nu + \delta\nu(t,r,\chi,\theta,\phi),  
\label{ptr-nu}  \\  
&&g_{\mu\nu} = \bar g_{\mu\nu} + h_{\mu\nu},
\label{ptrg} \\
&& T = 0 + t.
\label{ptr-t}
\end{eqnarray}
Here $h_{\mu\nu}$ is given by
\begin{equation}
h^\mu_{~\nu} = 
\left (
\begin{array}{ccccc}
h_1& h_3 & 0 & 0 & 0 \\
-d^2 h_3 /f & h_2 & 0 & 0 & 0 \\
0 & 0 & h^\chi_{~\chi} & h^\chi_{~\theta} & h^\chi_{~\phi} \\
0 & 0 & h^\theta_{~\chi} & h^\theta_{~\theta} & h^\theta_{~\phi} \\
0 & 0 & h^\phi_{~\chi} & h^\phi_{~\theta} & h^\phi_{~\phi} 
\end{array}
\right )
\label{ptr-h}
\end{equation}
This seems to be general for the s-wave calculation.

One has to linearize (\ref{eq-ricci})-(\ref{eq-t}) in order to obtain 
the equations governing the perturbations as
\begin{eqnarray}
&&  \delta R_{\mu\nu} (h)  
-{4 \over 3}(\partial_\mu \bar \lambda \partial_\nu \delta \lambda +
\partial_\mu \delta \lambda \partial_\nu \bar \lambda)
-4(\partial_\mu \bar \nu \partial_\nu \delta \nu +
\partial_\mu \delta \nu \partial_\nu \bar \nu) \nonumber \\
&&~~~~+{1 \over 2} e^{{8 \over 3} \bar \lambda} 
\bar F^{(K)}_{\mu \rho} \bar F^{(K)}_{\nu \alpha}h^{ \rho\alpha}   
-e^{{8 \over 3} \bar \lambda}\bar F^{(K)}_{\mu \rho} {\cal F}^{(K)~\rho}_{~~~~\nu} 
-{4 \over 3}e^{{8 \over 3} \bar \lambda}
\bar F^{(K)}_{\mu \rho} \bar F^{(K)~\rho}_{~~~~\nu} \delta\lambda 
\nonumber  \\
&&~~~~
+{1 \over 6} e^{{8 \over 3} \bar \lambda} 
\bar F^{(K)}_{\rho\sigma} {\cal F}^{(K)\rho\sigma}\bar g_{\mu\nu}   
-{1 \over 6}e^{{8 \over 3} \bar \lambda}\bar F^{(K)}_{\rho\kappa} 
\bar F^{(K)~\kappa}_{~~~~\eta} h^{\rho\eta} \bar g_{\mu\nu}
+{2 \over 9}e^{{8 \over 3} \bar \lambda}
\bar F^{(K)2}\bar g_{\mu\nu} \delta\lambda 
+{1 \over 12}e^{{8 \over 3} \bar \lambda} \bar F^{(K)2} h_{\mu\nu} 
\nonumber  \\
&&~~~~
+{1 \over 2} e^{-{4 \over 3} \bar \lambda + 4 \bar \nu} 
\bar F_{+\mu \rho} \bar F_{+\nu \alpha}h^{ \rho\alpha}   
-e^{-{4 \over 3} \bar \lambda + 4 \bar \nu}
\bar F_{+\mu \rho} {\cal F}^{~~\rho}_{+\nu} 
-e^{-{4 \over 3} \bar \lambda+4 \bar \nu}
\bar F_{+\mu \rho} \bar F^{~~\rho}_{+\nu} 
\left (-{2 \over 3}\delta\lambda +2 \delta \nu \right )
\nonumber  \\
&&~~~~
+{1 \over 6} e^{-{4 \over 3} \bar \lambda+4 \bar \nu} 
\bar F_{+\rho\sigma} {\cal F}_+^{\rho\sigma}\bar g_{\mu\nu}   
-{1 \over 6}e^{-{4 \over 3} \bar \lambda +4 \bar \nu}\bar F_{+\rho\kappa} 
\bar F^{~~\kappa}_{+\eta} h^{\rho\eta} \bar g_{\mu\nu}
\nonumber  \\
&&~~~~
+e^{-{4 \over 3} \bar \lambda +4 \bar \nu}
\bar F_+^2\bar g_{\mu\nu} \left (-{1 \over 9}\delta\lambda + 
{1 \over 3} \delta \nu \right )
+{1 \over 12}e^{-{4 \over 3} \bar \lambda + 4 \bar \nu} \bar F_+^2 h_{\mu\nu} 
\nonumber  \\
&&~~~~
+{1 \over 2} e^{-{4 \over 3} \bar \lambda + 4 \bar \nu} 
\bar F_{-\mu \rho} \bar F_{-\nu \alpha}h^{ \rho\alpha}   
-e^{-{4 \over 3} \bar \lambda + 4 \bar \nu}
\bar F_{-\mu \rho} {\cal F}^{~~\rho}_{-\nu} 
-e^{-{4 \over 3} \bar \lambda+4 \bar \nu}
\bar F_{-\mu \rho} \bar F^{~~\rho}_{-\nu} 
\left (-{2 \over 3}\delta\lambda +2 \delta \nu \right )
\nonumber  \\
&&~~~~
+{1 \over 6} e^{-{4 \over 3} \bar \lambda+4 \bar \nu} 
\bar F_{-\rho\sigma} {\cal F}_-^{\rho\sigma}\bar g_{\mu\nu}   
-{1 \over 6}e^{-{4 \over 3} \bar \lambda +4 \bar \nu}\bar F_{-\rho\kappa} 
\bar F^{~~\kappa}_{-\eta} h^{\rho\eta} \bar g_{\mu\nu}
\nonumber  \\
&&~~~~
+e^{-{4 \over 3} \bar \lambda +4 \bar \nu}
\bar F_-^2\bar g_{\mu\nu} \left (-{1 \over 9}\delta\lambda + 
{1 \over 3} \delta \nu \right )
+{1 \over 12}e^{-{4 \over 3} \bar \lambda + 4 \bar \nu} \bar F_-^2 h_{\mu\nu} 
\nonumber  \\
&&~~~~
+{1 \over 2} e^{-{4 \over 3} \bar \lambda - 4 \bar \nu} 
\bar H_{+\mu \rho} \bar H_{+\nu \alpha}h^{ \rho\alpha}   
-e^{-{4 \over 3} \bar \lambda - 4 \bar \nu}
\bar H_{+\mu \rho} {\cal H}^{~~\rho}_{+\nu} 
+e^{-{4 \over 3} \bar \lambda-4 \bar \nu}
\bar H_{+\mu \rho} \bar H^{~~\rho}_{+\nu} 
\left ({2 \over 3}\delta\lambda +2 \delta \nu \right )
\nonumber  \\
&&~~~~+{1 \over 6} e^{-{4 \over 3} \bar \lambda-4 \bar \nu} 
\bar H_{+\rho\sigma} {\cal H}_+^{\rho\sigma}\bar g_{\mu\nu}   
-{1 \over 6}e^{-{4 \over 3} \bar \lambda -4 \bar \nu}\bar H_{+\rho\kappa} 
\bar H^{~~\kappa}_{+\eta} h^{\rho\eta} \bar g_{\mu\nu}
\nonumber  \\
&&~~~~-e^{-{4 \over 3} \bar \lambda -4 \bar \nu}
\bar H_+^2\bar g_{\mu\nu} \left ({1 \over 9}\delta\lambda + 
{1 \over 3} \delta \nu \right )
+{1 \over 12}e^{-{4 \over 3} \bar \lambda - 4 \bar \nu} \bar H_+^2 h_{\mu\nu} 
\nonumber  \\
&&~~~~
+{1 \over 2} e^{-{4 \over 3} \bar \lambda - 4 \bar \nu} 
\bar H_{-\mu \rho} \bar H_{-\nu \alpha}h^{ \rho\alpha}   
-e^{-{4 \over 3} \bar \lambda - 4 \bar \nu}
\bar H_{-\mu \rho} {\cal H}^{~~\rho}_{-\nu} 
+e^{-{4 \over 3} \bar \lambda-4 \bar \nu}
\bar H_{-\mu \rho} \bar H^{~~\rho}_{-\nu} 
\left ({2 \over 3}\delta\lambda +2 \delta \nu \right )
\nonumber  \\
&&~~~~+{1 \over 6} e^{-{4 \over 3} \bar \lambda-4 \bar \nu} 
\bar H_{-\rho\sigma} {\cal H}_-^{\rho\sigma}\bar g_{\mu\nu}   
-{1 \over 6}e^{-{4 \over 3} \bar \lambda -4 \bar \nu}\bar H_{-\rho\kappa} 
\bar H^{~~\kappa}_{-\eta} h^{\rho\eta} \bar g_{\mu\nu}
\nonumber  \\
&&~~~~-e^{-{4 \over 3} \bar \lambda -4 \bar \nu}
\bar H_-^2\bar g_{\mu\nu} \left ({1 \over 9}\delta\lambda + 
{1 \over 3} \delta \nu \right )
+{1 \over 12}e^{-{4 \over 3} \bar \lambda - 4 \bar \nu} \bar H_-^2 h_{\mu\nu} 
=0,
\label{linR} \\
&&  \bar \nabla^2 \delta \nu
- h^{\mu\nu} \bar \nabla_\mu \bar \nabla_\nu \bar \nu   
- \bar g^{\mu\nu} \delta \Gamma^\rho_{\mu\nu} (h) \partial_\rho \bar \nu
\nonumber \\
&&~~~~~~-{1 \over 4} e^{-{4 \over 3} \bar \lambda+4 \bar \nu}
\bar F_{+\mu \nu} {\cal F}_+^{\mu \nu}                
+{1 \over 4} e^{-{4 \over 3} \bar \lambda+4 \bar \nu} 
\bar F_{+\mu \nu}   \bar F_{+\rho}^{~~\nu} h^{\mu\rho} 
-{1 \over 8}e^{-{4 \over 3} \bar \lambda+4 \bar \nu} \bar F_+^2 
\left ( -{4 \over 3}\delta \lambda + 4 \delta \nu \right )  \nonumber  \\
&&~~~~~~-{1 \over 4} e^{-{4 \over 3} \bar \lambda+4 \bar \nu}
\bar F_{-\mu \nu} {\cal F}_-^{\mu \nu}                
+{1 \over 4} e^{-{4 \over 3} \bar \lambda+4 \bar \nu} 
\bar F_{-\mu \nu}   \bar F_{-\rho}^{~~\nu} h^{\mu\rho} 
-{1 \over 8}e^{-{4 \over 3} \bar \lambda+4 \bar \nu} \bar F_-^2 
\left ( -{4 \over 3}\delta \lambda + 4 \delta \nu \right )  \nonumber  \\
&&~~~~~~+{1 \over 4} e^{-{4 \over 3} \bar \lambda-4 \bar \nu}
\bar H_{+\mu \nu} {\cal H}_+^{\mu \nu}                
-{1 \over 4} e^{-{4 \over 3} \bar \lambda-4 \bar \nu} 
\bar H_{+\mu \nu}   \bar H_{+\rho}^{~\nu} h^{\mu\rho} 
-{1 \over 8}e^{-{4 \over 3} \bar \lambda-4 \bar \nu} \bar H_+^2 
\left ( {4 \over 3}\delta \lambda + 4 \delta \nu \right )   \nonumber \\
&&~~~~~~+{1 \over 4} e^{-{4 \over 3} \bar \lambda-4 \bar \nu}
\bar H_{-\mu \nu} {\cal H}_-^{\mu \nu}                
-{1 \over 4} e^{-{4 \over 3} \bar \lambda-4 \bar \nu} 
\bar H_{-\mu \nu}   \bar H_{-\rho}^{~\nu} h^{\mu\rho} 
-{1 \over 8}e^{-{4 \over 3} \bar \lambda-4 \bar \nu} \bar H_-^2 
\left ( {4 \over 3}\delta \lambda + 4 \delta \nu \right )   = 0,
\label{lin-nu}  \\
&&  \bar \nabla^2 \delta \lambda
- h^{\mu\nu} \bar \nabla_\mu \bar \nabla_\nu \bar \lambda   
- \bar g^{\mu\nu} \delta \Gamma^\rho_{\mu\nu} (h) \partial_\rho \bar \lambda
\nonumber \\
&&~~~~~~-{1 \over 2} e^{{8 \over 3} \bar \lambda}
\bar F^{(K)}_{\mu \nu} {\cal F}^{(K)\mu \nu}                
+{1 \over 2} e^{{8 \over 3} \bar \lambda} 
\bar F^{(K)}_{\mu \nu}   \bar F_{~~~~\rho}^{(K)~\nu} h^{\mu\rho} 
-{2 \over 3} e^{{8 \over 3} \bar \lambda} \bar F^{(K)2} 
\delta \lambda  \nonumber  \\
&&~~~~~~+{1 \over 4} e^{-{4 \over 3} \bar \lambda+4 \bar \nu}
\bar F_{+\mu \nu} {\cal F}_+^{\mu \nu}                
-{1 \over 4} e^{-{4 \over 3} \bar \lambda+4 \bar \nu} 
\bar F_{+\mu \nu}   \bar F_{+\rho}^{~~\nu} h^{\mu\rho} 
+e^{-{4 \over 3} \bar \lambda+4 \bar \nu} \bar F_+^2 
\left ( -{1 \over 6}\delta \lambda + {1 \over 2} \delta \nu \right )  
\nonumber  \\
&&~~~~~~+{1 \over 4} e^{-{4 \over 3} \bar \lambda+4 \bar \nu}
\bar F_{-\mu \nu} {\cal F}_-^{\mu \nu}                
-{1 \over 4} e^{-{4 \over 3} \bar \lambda+4 \bar \nu} 
\bar F_{-\mu \nu}   \bar F_{-\rho}^{~~\nu} h^{\mu\rho} 
+e^{-{4 \over 3} \bar \lambda+4 \bar \nu} \bar F_-^2 
\left ( -{1 \over 6}\delta \lambda + {1 \over 2} \delta \nu \right )  
\nonumber  \\
&&~~~~~~+{1 \over 4} e^{-{4 \over 3} \bar \lambda-4 \bar \nu}
\bar H_{+\mu \nu} {\cal H}_+^{\mu \nu}                
-{1 \over 4} e^{-{4 \over 3} \bar \lambda-4 \bar \nu} 
\bar H_{+\mu \nu}   \bar H_{+\rho}^{~~\nu} h^{\mu\rho} 
-e^{-{4 \over 3} \bar \lambda-4 \bar \nu} \bar H_+^2 
\left ( {1 \over 6}\delta \lambda + {1 \over 2} \delta \nu \right )  
\nonumber \\
&&~~~~~~+{1 \over 4} e^{-{4 \over 3} \bar \lambda-4 \bar \nu}
\bar H_{-\mu \nu} {\cal H}_-^{\mu \nu}                
-{1 \over 4} e^{-{4 \over 3} \bar \lambda-4 \bar \nu} 
\bar H_{-\mu \nu}   \bar H_{-\rho}^{~~\nu} h^{\mu\rho} 
-e^{-{4 \over 3} \bar \lambda-4 \bar \nu} \bar H_-^2 
\left ( {1 \over 6}\delta \lambda + {1 \over 2} \delta \nu \right )   = 0,
\label{lin-lambda}  \\
&&\left ( \bar \nabla_\mu +{8 \over 3} \partial_\mu \bar \lambda \right ) 
    \left ( {\cal F}^{(K)\mu \nu} - \bar F_{~~~~\alpha}^{(K)~\nu} h^{\alpha \mu}
                    - \bar F^{(K)\mu}_{~~~~~\beta} h^{\beta \nu} \right )
      + \bar F^{(K)\mu \nu} \left (\delta \Gamma^\sigma_{\sigma \mu} (h)       
      +{8 \over 3} \partial_\mu \delta \lambda \right ) 
 = 0,  \label{linFk} \\
&&\left (\bar\nabla_\mu -{4 \over 3}\partial_\mu \bar\lambda+4 \partial_\mu \bar \nu \right ) 
    \left ( {\cal F}_+^{\mu \nu} - \bar F_{+\alpha}^{~~\nu} h^{\alpha \mu}
                             - \bar F^{\mu}_{+\beta} h^{\beta \nu} \right )
\nonumber \\
&&\qquad \qquad
      + \bar F_+^{\mu \nu} \left (\delta \Gamma^\sigma_{\sigma \mu} (h)       
      -{4 \over 3} \partial_\mu \delta \lambda 
      +4 \partial_\mu \delta \nu + \partial_\mu t \right ) 
 = 0,  \label{linFp} \\
&&\left (\bar\nabla_\mu -{4 \over 3}\partial_\mu \bar\lambda+4 \partial_\mu \bar \nu \right ) 
    \left ( {\cal F}_-^{\mu \nu} - \bar F_{-\alpha}^{~~\nu} h^{\alpha \mu}
                             - \bar F^{\mu}_{-\beta} h^{\beta \nu} \right )
\nonumber \\
&&\qquad \qquad
      + \bar F_-^{\mu \nu} \left (\delta \Gamma^\sigma_{\sigma \mu} (h)       
      -{4 \over 3} \partial_\mu \delta \lambda 
      +4 \partial_\mu \delta \nu - \partial_\mu t \right ) 
 = 0,  \label{linFm} \\
&&\left (\bar\nabla_\mu -{4 \over 3}\partial_\mu \bar\lambda-4 \partial_\mu \bar \nu \right ) 
    \left ( {\cal H}_+^{\mu \nu} - \bar H_{+\alpha}^{~~\nu} h^{\alpha \mu}
                             - \bar H^{\mu}_{+\beta} h^{\beta \nu} \right )
\nonumber \\
&&\qquad \qquad
      + \bar H_+^{\mu \nu} \left (\delta \Gamma^\sigma_{\sigma \mu} (h)       
      -{4 \over 3} \partial_\mu \delta \lambda 
      -4 \partial_\mu \delta \nu + \partial_\mu t \right ) 
 = 0,  \label{linHp} \\
&&\left (\bar\nabla_\mu -{4 \over 3}\partial_\mu \bar\lambda-4 \partial_\mu \bar \nu \right ) 
    \left ( {\cal H}_-^{\mu \nu} - \bar H_{-\alpha}^{~~\nu} h^{\alpha \mu}
                             - \bar H^{\mu}_{-\beta} h^{\beta \nu} \right )
\nonumber \\
&&\qquad \qquad
      + \bar H_-^{\mu \nu} \left (\delta \Gamma^\sigma_{\sigma \mu} (h)       
      -{4 \over 3} \partial_\mu \delta \lambda 
      -4 \partial_\mu \delta \nu - \partial_\mu t \right ) 
 = 0,  \label{linHm} \\
&& \left (\bar \nabla^2 - m^2 e^{-2 \bar \lambda /3} \right ) t 
 - e^{- 4 \bar \lambda /3 + 4 \bar \nu} 
     \left \{\bar F_+^2 \left ( {t\over 2}+{\cal F}_+ \right )+ 
\bar F_-^2 \left ( {t\over 2}-{\cal F}_- \right ) \right \}
\nonumber \\
&&\qquad\qquad\qquad
 - e^{- 4 \bar \lambda /3 - 4 \bar \nu}
     \left \{\bar H_+^2 \left ( {t \over 2}+{\cal H}_+ \right )+ 
     \bar H_-^2 \left ( {t\over 2}-{\cal H}_- \right ) \right \}
=0, \label{lint}
\end{eqnarray}
where 
\begin{eqnarray}
&&\delta R_{\mu\nu} (h) = 
 - {1 \over 2} \bar \nabla^2 h_{\mu\nu}   
 -{1 \over 2} \bar \nabla_\mu \bar \nabla_\nu h^\rho_{~\rho}
 + {1 \over 2} \bar \nabla^\rho \bar \nabla_\nu h_{\rho\mu}   
 + {1 \over 2} \bar \nabla^\rho \bar \nabla_\mu h_{\nu\rho},   
\label{delR} \\   
&&\delta \Gamma^\rho_{\mu\nu} (h) = {1 \over 2} \bar g^{\rho\sigma} 
( \bar \nabla_\nu h_{\mu\sigma} + \bar \nabla_\mu h_{\nu\sigma} - \bar \nabla_\sigma h_{\mu\nu} ).
\label{delGam}
\end{eqnarray}
Since we start with full degrees of freedom (\ref{ptr-h}), we 
choose a gauge to study the propagation of fields.
For this purpose $\delta R_{\mu\nu}$ can be transformed into the 
Lichnerowicz operator\cite{Gre94NPB399}
\begin{equation}
\delta R_{\mu\nu} = - {1 \over 2} {\bar \nabla}^2 h_{\mu\nu} 
+ {\bar R}_{\sigma ( \nu} h^\sigma_{~\mu)}
- {\bar R}_{\sigma \mu \rho \nu} h^{\sigma \rho} 
+ {\bar \nabla}_{(\nu} {\bar \nabla}_{|\rho|} {\hat h}^\rho_{~\mu)}.
\label{lichnerowicz}
\end{equation}
We have to examine whether there exists any choice of gauge which can simplify 
Eqs.(\ref{lin-nu}) and (\ref{lin-lambda}).
Conventionally we choose the harmonic (transverse) gauge
(${\bar \nabla}_\mu {\hat h}^{\mu\rho} = \bar g^{\mu\nu} 
\delta \Gamma^\rho_{\mu\nu} = 0 $).

Considering the harmonic gauge and $Q_1=Q_5$ case for simplicity, 
Eqs.(\ref{lin-nu}),  
(\ref{lin-lambda}) and (\ref{lint}) lead to
\begin{eqnarray}
&&{\bar \nabla}^2 \delta \nu + { Q_1^2 \over r^6 f_1^2 f^{1/3}}
  \left (2 {\cal F}_+ + 2 {\cal F}_- -2 {\cal H}_+ -2 {\cal H}_- + 
  16 \delta \nu \right ) =0, 
\label{nu-harmonic} \\
&&{\bar \nabla}^2 \delta \lambda  
- { d \over f^{1/3} } h^{rr} \partial^2_r \bar \lambda
+ { d \over f^{1/3}} h^{\mu\nu} \Gamma^r_{\mu\nu} \partial_r \bar \lambda
- {2 \tilde Q_K^2 \over r^6 f_K^2 f^{1/3}} \left ( h_1 + h_2 - 2 {\cal F}^{(K)} 
          -{8 \over 3} \delta \lambda \right ) 
\nonumber \\
&&~~~~~~+ { 2 Q_1^2 \over r^6 f_1^2 f^{1/3}}
  \left ( 2 h_1 +  2 h_2 -  {\cal F}_+ -  {\cal F}_- 
- {\cal H}_+ - {\cal H}_- + {8 \over 3}\delta \lambda \right ) =0, 
\label{lambda-harmonic}  \\
&& \bar \nabla^2 t + { 2 \over \alpha'} 
\left ( { f_1 \over f_K} \right )^{1/3}  t 
+ {{8 Q_1^2} \over {r^6 f_1^2 f^{1/3}}} 
\left ( 2 t + {\cal F}_+ - {\cal F}_- + {\cal H}_+ - {\cal H}_- \right )
=0, \label{t-harmonic}
\end{eqnarray}
Now we attempt to disentangle the mixing between 
($\delta \nu, \delta \lambda, t$)
and other fields by using both the harmonic gauge and U(1) 
field equations in Eqs.(\ref{linFk})-(\ref{linHm}).
After some calculations, one finds the relations 
\begin{eqnarray}
2 {\cal F}^{(K)} &=& h_1 + h_2 - h^{\theta_i}_{~\theta_i} 
            - { 16 \over 3} \delta \lambda,
\label{solFk} \\
2 ({\cal F}_\pm \pm t)  &=& h_1 + h_2 - h^{\theta_i}_{~\theta_i} 
            + { 8 \over 3} \delta \lambda - 8 \delta \nu,
\label{solF} \\
2 ({\cal H}_\pm \pm t) &=& h_1 + h_2 - h^{\theta_i}_{~\theta_i} 
            + { 8 \over 3} \delta \lambda + 8 \delta \nu,
\label{solH}
\end{eqnarray}
where $h^{\theta_i}_{~\theta_i} = h^\chi_{~\chi} +  h^\theta_{~\theta} +
 h^\phi_{~\phi}$.
Using (\ref{solFk})-(\ref{solH}), one obtains the equations
for $\delta \nu$, $\delta \lambda$ and $t$  as
\begin{eqnarray}
&&{\bar \nabla}^2 \delta \nu -{8 \tilde Q_1^2 \over r^6 f_1^2 f^{1/3}} \delta \nu =0, 
\label{nu-decoupled} \\
&&{\bar \nabla}^2 \delta \lambda  
- { d \over f^{1/3} } h^{rr} \partial^2_r \bar \lambda
+ { d \over f^{1/3}} h^{\mu\nu} \Gamma^r_{\mu\nu} \partial_r \bar \lambda
+ {2 \over r^6 f^{1/3}} \left ( {\tilde Q_1^2 \over f_1^2} 
    - {\tilde Q_K^2 \over f_K^2} \right ) h^{\theta_i}_{~\theta_i} 
\nonumber \\
&&~~~~~~
- {8 \over 3 r^6 f^{1/3}} \left ( {\tilde Q_1^2 \over f_1^2} 
    + 2 {\tilde Q_K^2 \over f_K^2} \right ) \delta \lambda =0,
\label{lambda-decoupled} \\
&&
\bar \nabla^2 t + \left \{ 2 \left ( {f_1 \over f_K} \right )^{1/3} 
  - {{8 \tilde Q_1^2} \over {r^6 f_1^2 f^{1/3}}} \right \} t =0.
\label{t-decoupled}
\end{eqnarray}
We wish to point out that $(\delta \nu, t)$-equations are decoupled 
completely but 
$\delta \lambda$-equation still remains a coupled form.
In this sense the role of $\delta \lambda$ remains obscure.
Hence we no longer consider this field.
Here we observe from (\ref{nu-decoupled}) and (\ref{t-decoupled}) that 
if the mass term in (\ref{t-decoupled}) is absent, two equations 
are exactly the same form.

\section{Potential Analysis}
\label{potential}
From the Bianchi identities (\ref{bianchi}) one has 
\begin{eqnarray}
&&\partial_\chi {\cal F}^{(K)} = \partial_\theta {\cal F}^{(K)} =
      \partial_\phi {\cal F}^{(K)}=0, \nonumber \\
&&\partial_\chi {\cal F}_\pm = \partial_\theta {\cal F}_\pm =
      \partial_\phi {\cal F}_\pm=0, \nonumber \\
&&\partial_\chi {\cal H}_\pm = \partial_\theta {\cal H}_\pm =
      \partial_\phi {\cal H}_\pm=0. \label{eq-bianchi} 
\end{eqnarray}
This implies either ${\cal F}^{(K)}={\cal F}^{(K)} (t,r), {\cal F}_\pm=
{\cal F}_\pm(t,r), {\cal H}_\pm={\cal H}_\pm(t,r)$ or 
${\cal F}^{(K)}={\cal F}_\pm= {\cal H}_\pm=0$.  
The former together with (\ref{solFk})-(\ref{solH}) means 
that all higher modes of $l \ge 1$ are forbidden in this scheme.  
Hence we consider only the s-wave($l$=0) propagations.
Then the relevant fields become 
$\delta \nu(r,t) = \delta \nu(r) e^{i \omega t}$ 
and $t(r,t) = t(r) e^{i \omega t}$, 
but the graviton modes $h_{\mu\nu}$ are irrelevant to our interest.
For $r_1=r_5\equiv R$, the equations (\ref{nu-decoupled}) and 
(\ref{t-decoupled}) lead to 
\begin{eqnarray}
&&\left \{ r^{-3} \partial_r (d r^3 \partial_r) - d^{-1} f \partial^2_t 
-{{8 R^4} \over {r^2(r^2 + R^2)^2}}\left ( 1 + {r_0^2 \over R^2} 
\right ) \right \} \delta \nu  =0,
\label{reduce-eqnu}\\
&&\left \{ r^{-3} \partial_r (d r^3 \partial_r) - d^{-1} f \partial^2_t 
+ { 2 \over {\left (1+{r_K^2 \over r^2} \right )^{2/3} 
\left (1 + {R^2 \over r^2 } \right )^{1/3}}}
-{{8 R^4} \over {r^2(r^2 + R^2)^2}}\left ( 1 + {r_0^2 \over R^2} 
\right ) \right \} t =0,
\label{reduce-eqt}
\end{eqnarray}
Considering $ N = r^{-3/2} \tilde N$,
for $N = \delta \nu, t$ and
introducing a tortoise coordinate
$ r^* = \int {(dr/d(r))} = r + (r_0/2)\ln |(r - r_0)
/(r + r_0)|$\cite{Lee9903054}, then the equation takes the form
\begin{equation}
{d^2 \tilde N \over d r^{*2} }
+ (\omega^2 - \tilde V_N) \tilde N = 0.
\label{eqom}
\end{equation}
Here we take $r_0 = r_K$ for simplicity.
In the dilute gas limit $(R \gg r_0), \tilde V_N(r)$ is given by
\begin{equation}
\tilde V_\nu(r) = - \omega^2 (f -1)
+ h \left \{ { 3   \over 4 r^2 } \left (1 + { 3 r_0^2 \over r^2} \right ) +
{ 8 R^4  \over r^2 (r^2 + R^2)^2 } \right \},
\label{potential-nu}
\end{equation}
\begin{equation}
\tilde V_t(r) = - \omega^2 (f -1)
+ h \left \{ { 3   \over 4 r^2 } \left (1 + { 3 r_0^2 \over r^2} \right ) +
{ 8 R^4  \over r^2 (r^2 + R^2)^2 } 
- { 2 \over {(1 + r_0^2/r^2)^{2/3} (1 + R^2/r^2)^{1/3}}}
\right \},
\label{potential-t}
\end{equation}
where
\begin{equation}
f -1 = { r_0^2 + 2 R^2 \over r^2} + { (2 r^2_0 + R^2) R^2 \over r^4}
        + { r^2_0 R^4 \over r^6}.
\end{equation}
\begin{figure}
\epsfig{file=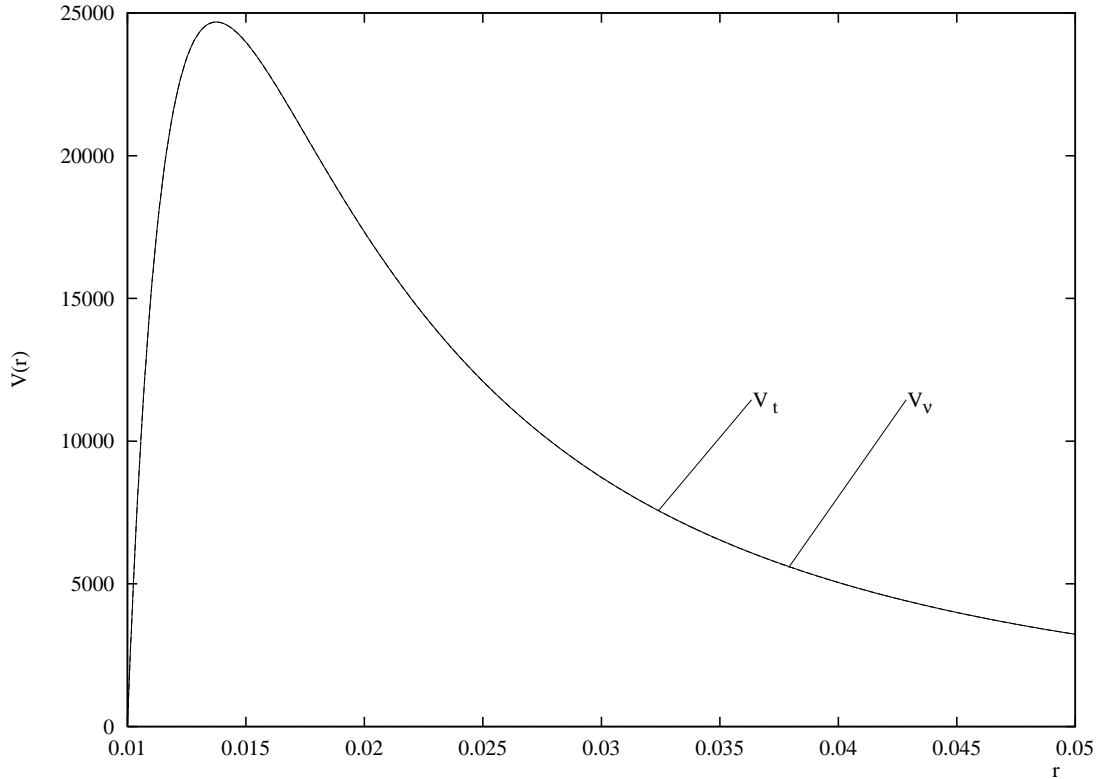,width=0.9\textwidth,clip=}
\label{ij9904-near}
\caption{
The graph of $V_\nu(r), V_t(r)$ in the near-horizon region with $r_0=0.01, R=0.3$.
}
\end{figure}

We note that $\tilde V_N$ depends on two parameters ($r_0, R$)
as well as the energy$(\omega)$. As (\ref{eqom}) stands, it is far from
the Schr\"odinger-type equation. The $\omega$-dependence is
a matter of peculiar interest
to us compared with the Schwarzschild black hole
potentials $(V_{RW}, V_Z, V_\psi)$\cite{Cha83,Kwo86PRD333}. This makes the
interpretation of $\tilde V_N$ as a potential difficult.
This arises
because $(f-1)$ is very large as $10^6$ for $r_0 = 0.01,
R= 0.3 $ in the near-horizon.
In order for $\tilde V_N$ to be a potential,
it is necessary to take the low energy limit of $\omega \to 0$.
It is suitable to be $10^{-3}$.
And $\omega^2(f-1)$ is of order ${\cal O}(1)$ and thus
it can be ignored in comparison with the remaing ones.
Now we can define a potential
$V_N = \tilde V_N + \omega^2 ( f- 1)$.
Hence, in the low energy limit($\omega \to 0$), Eq.(\ref{eqom})
becomes as the Schr\"odinger-type equation.

First we consider the near-horizon geometry, which corresponds to the 
limit of $r \to r_0$ such that the dilute gas limit
($R \gg r_0$) holds.
In this limit one finds AdS$_3 \times$S$^3 \times$T$^4$. 
In the near-horizon, as is shown in Fig. 1, 
the potential of the fixed scalar($V_\nu(r)$) takes exactly the 
same form of the tachyon($V_t(r)$).
This means that in the near-horizon the roles of the fixed 
scalar($\nu$) and the tachyon($T$) are the same.
Now let us observe the far-region behavior of their potentials.
In the asymptotic region($r \gg R$) $V_\nu$ approaches to zero,
whereas $V_t$ goes to $-2$(see Fig. 2). 
With $V_t$ one finds an exponentially growing mode
($e^{i \omega t}, \omega = -i \alpha$) 
from Eq. (\ref{eqom})\cite{Kwo86PRD333}.
This implies that $\nu$ plays a role of the good test field, while 
$t$ induces an instability of the flat space.

\begin{figure}
\epsfig{file=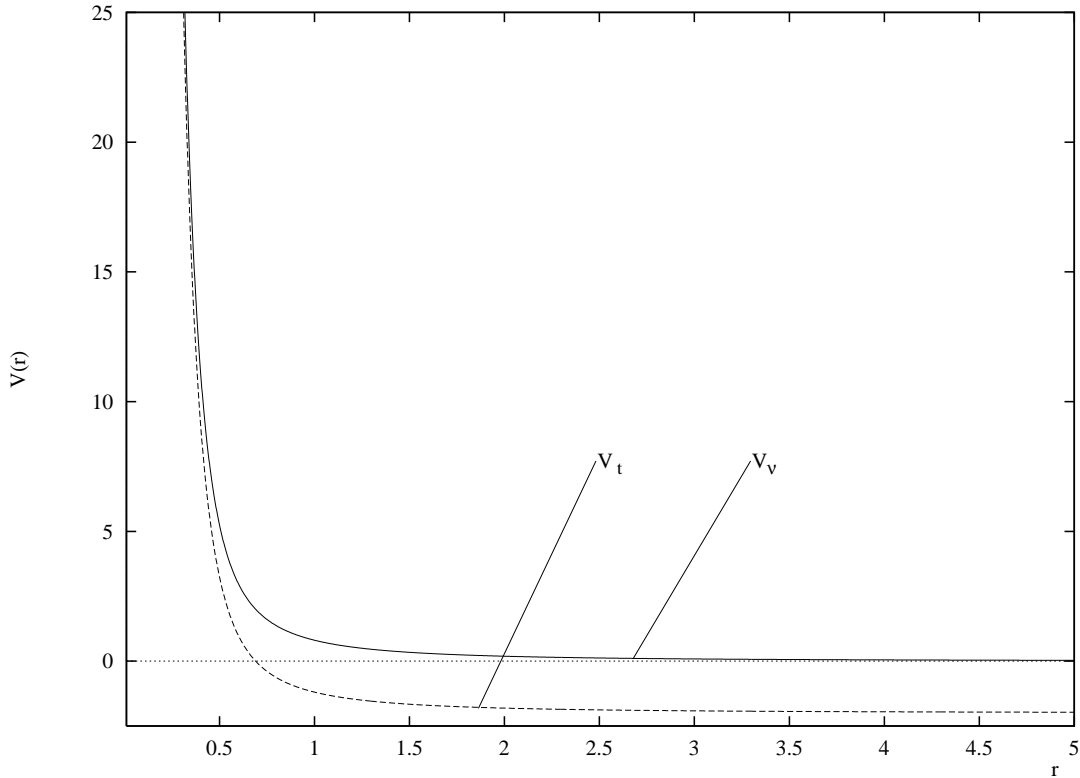,width=0.9\textwidth,clip=}
\label{ij9904-far}
\caption{
The graph of $V_\nu(r), V_t(r)$ in the far-region with $r_0=0.01, R=0.3$.
}
\end{figure}

\section{Discussions}
\label{discussion}
Let us first discuss the role of a fixed scalar $\nu$ in the 
greybody factor calculation(Hawking radiation).  Although $\nu$ is 
related to the scale of $T^4$, it turns out to be the 10D 
dilaton($\phi_{10}$) when $\phi_6 = \phi_{10} - 2 \nu =0$.  
For $Q_1=Q_5$ case, one finds the same linearized equation 
for the harmonic, dilaton gauge, and K-K setting
\cite{Lee98PRD104006}.  This means that the 
fixed scalar($\nu$) gives us a gauge-invariant result.  
In the low energy limit ($\omega \to 0$), 
the s-wave semiclassical greybody factor takes the form\cite{Lee9903054}
\begin{equation}
\sigma^\nu_{abs} = { {\cal A}_H \over 4 }  \left ( r_0 \over R \right ) ^4.
\label{abs-nu}
\end{equation}

On the other hand, $\lambda(=\nu_5 - \phi_6/2$) is entirely determined by  
the scale($\nu_5$) of the KK circle($S^1$) when $\phi_6$ is 
turned off.  The semiclassical result of its greybody factor takes the form
\begin{equation}
\sigma^\lambda_{abs} = { 9 \over 4} {\cal A}_H \left ( {r_0 \over R} \right )^4.
\label{abs-lambda}
\end{equation}
However, in the previous work\cite{Lee98PRD104006} we 
found out that $\lambda$ depends on the gauge choice.  
The fixed scalar $\nu$ is clearly understood as a good test 
field for studying the D5$_\pm$-D1$_\pm$ brane black hole.  
On the other hand, the role of $\lambda$ as a test field is obscure because 
it is a gauge-dependent field and gives rise to some 
disagreement for the cross section.
The tachyon plays the same role of $\nu$ in the near-horizon geometry.
But this induces an instability of Minkowski space(see Fig. 2).
Hence the tachyon seems not to be a good test field to investigate 
the quantum aspect of the D5$_\pm$-D1$_\pm$ brane black hole.

Finally let us comment on the stability problem of the 
near-horizon geometry. 
It is known that while the Minkowski vacuum is unstable in 
type 0 string theory, the AdS$_5 \times $S$^5$ with self-dual 
5-form flux should be a stable background for sufficiently small
radius\cite{Kle99JHEP03015,Cos9903128}.
The RR fields work to stabilize the tachyon in the near-horizon.
It is clear from Fig. 1 that the AdS$_3 \times$S$^3 \times$T$^4$ is stable 
because $V_\nu(r)$ and $V_t(r)$ take the shapes of the potential barrier.
From Eq.(\ref{t-harmonic}), if there do not exist the RR fields
(${\cal F}_\pm, {\cal H}_\pm$), one finds a potential well for 
the tachyon, which induces an instability in the near-horizon\cite{Cha83}.
However, thanks to the RR fields, one obtains a potential barrier.
This shows obviously that the RR fields can work to stabilize the tachyon 
in the near-horizon.

\section*{Acknowledgement}
This work was supported in part by the Basic Science Research Institute 
Program, Ministry of Education, Project Nos. BSRI-98-2413, BSRI-98-2441 
and a grant from Inje University.

\end{document}